\documentclass[
    ,final            
  ]
  {aipproc}

\layoutstyle{8x11double}

\newcommand{\apj}{{\it ApJ}}

\newcommand{\apjl}{{\it ApJ}}
\newcommand{\aap}{{\it A\&A}}

\newcommand{\aapr}{{\it A\&A~Rev}}
\newcommand{\mnras}{{\it MNRAS}}

\newcommand{\nat}{{\it Nature}}

\def\sedici {SGR\,1627--41}
\def\sgr {1E\,1547.0--5408}
\def\pdot {\dot P}

\begin{document}

\title{The Twin Magnetars: SGR 1627--41 and   1E 1547--5408}


\author{S. Mereghetti}{
  address={INAF-IASF Milano, v. E.Bassini 15, 20133 Milano, Italy}}

\author{A. Tiengo}{
  address={INAF-IASF Milano, v. E.Bassini 15, 20133 Milano, Italy}}

\author{P. Esposito}{
  address={INAF-IASF Milano, v. E.Bassini 15, 20133 Milano, Italy}
  ,altaddress={INFN-Pavia, via A.~Bassi 6, 27100 Pavia, Italy}}

\author{G. Vianello}{
  address={INAF-IASF Milano, v. E.Bassini 15, 20133 Milano, Italy}}

\author{A. De Luca}{
  address={INAF-IASF Milano, v. E.Bassini 15, 20133 Milano, Italy}
 ,altaddress={IUSS - Istituto Universitario di Studi Superiori,
 viale Lungo Ticino Sforza 56, 27100 Pavia, Italy}}

\author{D. G\"{o}tz}{
  address={CEA Saclay, DSM/Irfu/Service d'Astrophysique, Orme des Merisiers, B\^at. 709, 91191 Gif sur Yvette, France}
}

\author{G. Weidenspointner}{
  address={MPI f\"{u}r extraterrestrische Physik,
Giessenbachstrasse,  Postfach 1312, D-85741 Garching, Germany}
  ,altaddress={MPI Halbleiterlabor, Otto-Hahn-Ring 6, 81739 Muenchen, Germany }}

\author{A. von Kienlin}{
  address={MPI f\"{u}r extraterrestrische Physik,
Giessenbachstrasse,  Postfach 1312, D-85741 Garching, Germany} }

\author{G.L. Israel}{
  address={INAF - Osservatorio Astronomico di Roma, via Frascati 33, 00040 Monteporzio Catone, Italy}}

\author{L. Stella}{
  address={INAF - Osservatorio Astronomico di Roma, via Frascati 33, 00040 Monteporzio Catone, Italy} }

\author{N. Rea}{
  address={Astronomical Institute ``Anton Pannekoek'', University of Amsterdam, Kruislaan 403, 1098~SJ Amsterdam, The Netherlands}
}

\author{R. Turolla}{
  address={Dipartimento di Fisica, Universit\`a degli Studi di Padova, via F.~Marzolo 8, 35131 Padova, Italy}
}

\author{S. Zane}{
  address={MSSL, University College London, Holmbury St. Mary, Dorking, Surrey RH5 6NT, UK}
}

\begin{abstract}
 We report on recent results obtained thanks to Target of Opportunity
 observations of  the two galactic sources SGR 1627--41 and 1E
 1547--5408. These two transient sources present several
 similarities which support the   interpretation of Anomalous
 X--ray Pulsars and Soft Gamma-ray Repeaters as a single class of
 strongly magnetized neutron stars.
\end{abstract}

\maketitle


\section{Introduction}

During the last decade, mounting evidence has been found
supporting the idea  that Soft Gamma-ray Repeaters (SGRs) and
Anomalous X-ray Pulsars (AXPs) are isolated neutron stars with
peculiar properties resulting from the presence of an ultra strong
magnetic field, B$\sim$10$^{14}$--10$^{15}$ G. The magnetar model
\cite{tho95,tho96}, initially developed to explain the SGRs, has
been quite successfully applied to both classes of sources, which
indeed show many commonalities. Actually, the distinction between
SGRs and AXPs might well be only  a semantic heritage linked to
the way these objects were first observed: the SGRs, discovered as
sources of short ($<$1 s) repeating bursts of hard X-rays
\cite{lar86}  have persistent counterparts practically
indistinguishable from the AXPs, while several AXPs, discovered as
spinning down X-ray pulsars without evidence of companions stars
\cite{mer95}, have been observed to emit short bursts very similar
to those of the SGRs. For a recent review on AXPs and SGRs see
\cite{mer08}.

Here we report on new X--ray observations of two sources, \sedici\
and \sgr , obtained as Target of Opportunity requests  to study
their recent outbursts. The results give further evidence for  the
similarities between AXPs and SGRs. Among the small group of known
magnetars (about 15 sources) the two objects discussed here stand
out for having the shortest spin periods: 2.6 s for \sedici\ and
2.1 s for \sgr . Both sources are transients and lie in supernova
remnants. Curiously, they are also located almost in the same
direction of the Galactic plane, hence their "twins" nickname.

\begin{figure}
  \includegraphics[width=14cm]{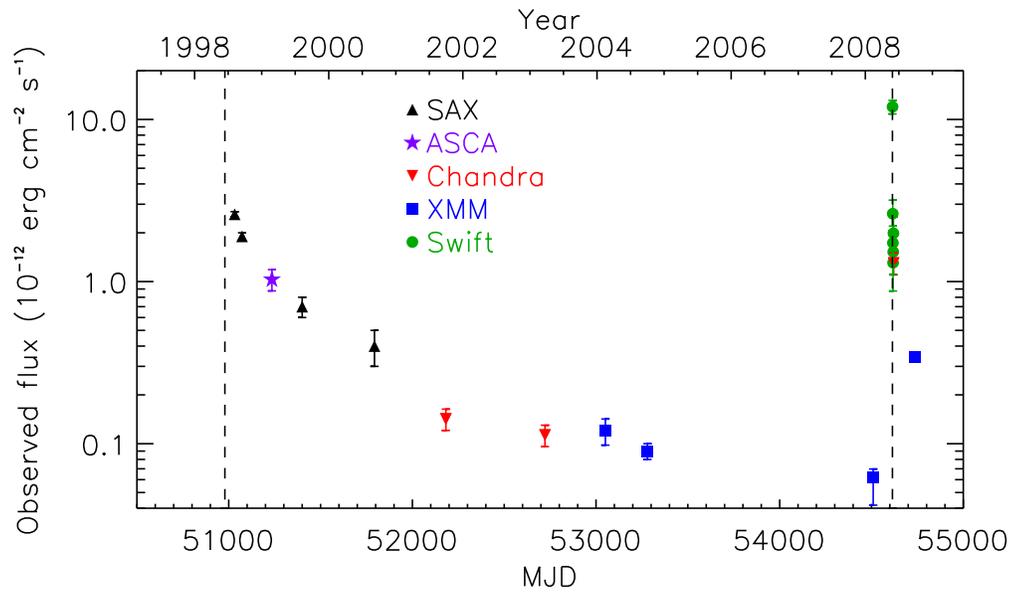}
 \caption{X--ray light curve of \sedici\   spanning ten years of
observations with different satellites (observed flux in the 2-10
keV energy range).  The vertical lines indicate the two periods of
bursting activity seen from this source (June 1998 and May 2008).
\label{fig:lc1627}}
\end{figure}

\begin{figure}
  \includegraphics[angle=270,width=10cm]{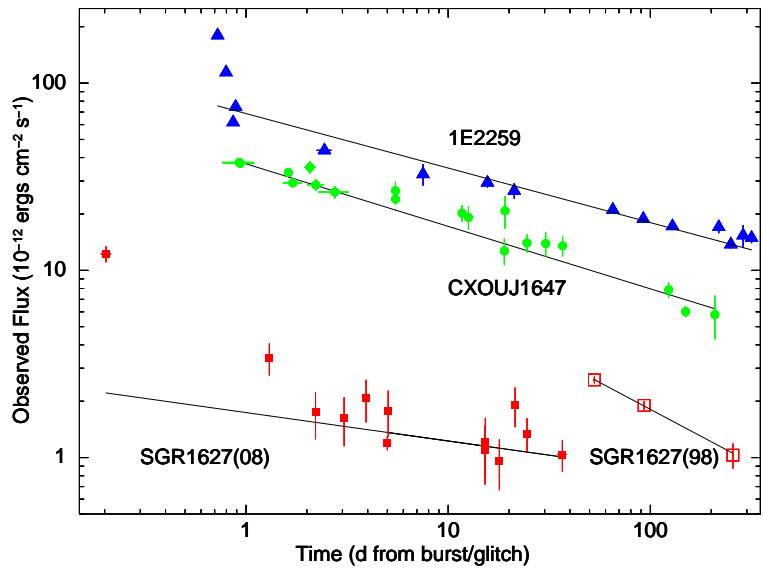}
 \caption{Comparison of the long term flux decays following outbursts
of SGRs and AXPs. 
For \sedici\ both the 1998 and 2008 events are plotted.
The lines are power laws with time decay index ranging from --0.2
(\sedici\ in 2008) to --0.6 (\sedici\ in 1998).
\label{fig:decays}}
\end{figure}

\section{SGR 1627--41}

\sedici\ was discovered in 1998, during a bursting state that
lasted about six weeks~\cite{woo99}. Its X-ray counterpart,
identified at the time of the outburst,  had a luminosity of $\sim
10^{35}$ erg s$^{-1}$ (for d=11 kpc). In the following years its
X--ray luminosity monotonically decreased~\cite{mer06a}, until the
lowest flux was observed in February 2008 with XMM-Newton (see
Fig.~\ref{fig:lc1627}). This flux corresponds to a luminosity of
only $\sim10^{33}$ erg s$^{-1}$, the lowest ever observed for a
SGR~\cite{esp08}. The long-term flux decay can be interpreted as
the cooling of the neutron star, assuming that the star was
significantly heated during the outburst. The modelling of the
long term light curve can provide information on the mechanism for
(and location of) the heating, as well as on the neutron star
structure~\cite{kou03}. However, this is complicated by the fact
that the measurements carried out over several years were obtained
with different satellites. This introduces some uncertainties due
to the relative calibrations. Furthermore, the emitted fluxes are
also difficult to estimate precisely, owing to the high absorption
of this source (several 10$^{22}$ cm$^{-2}$) and the limited
spectral information. In fact a reanalysis of all the available
data ~\cite{mer06a} showed that some features of the the long term
light curve, from which properties of the cooling model had been
inferred, are not statistically significant.

\begin{figure}
 \includegraphics[angle=270,width=7cm]{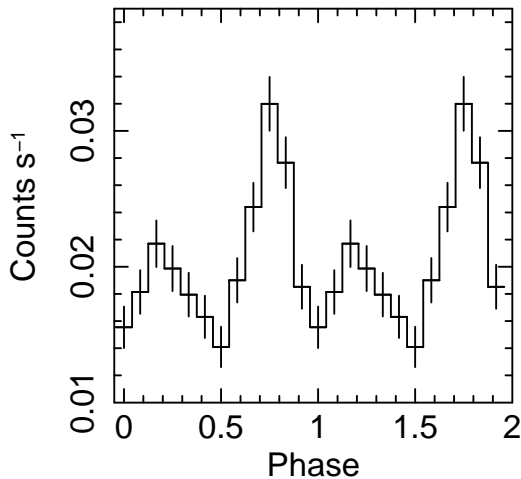}
 \caption{X-ray light curve of \sedici\ folded at the spin period of 2.59
s discovered by Esposito et al. (2009). The data have been
obtained with the XMM-Newton EPIC pn camera in the 2-12 keV energy
range. \label{fig:fol1627}}
\end{figure}

\begin{figure}
  \includegraphics[width=10cm]{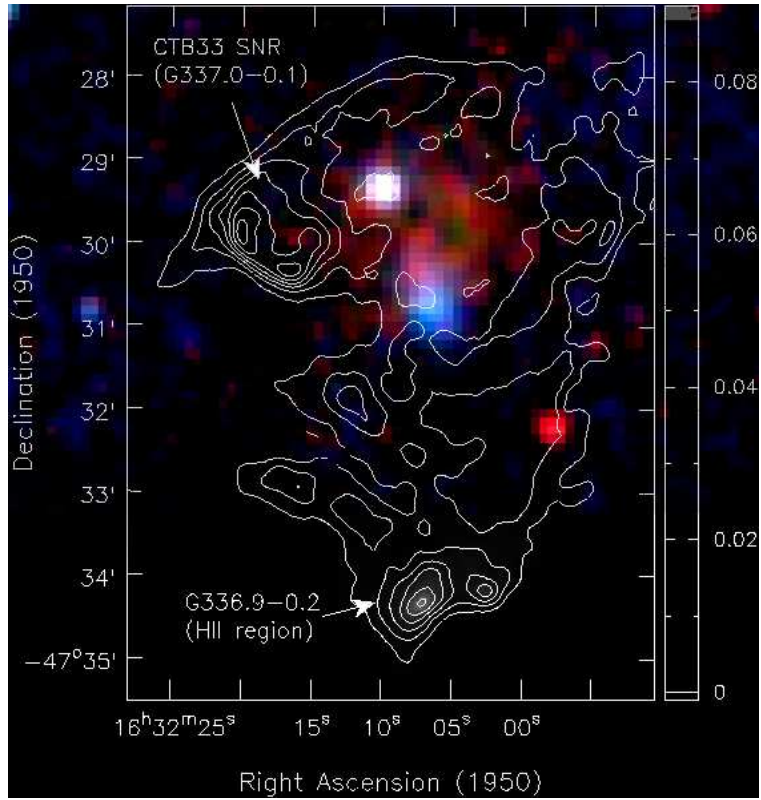}
\caption{XMM-Newton EPIC X-ray image of the region of \sedici\
with overlaid contours from the 1375 MHz radio map of Sarma et al.
(1997). The colors indicate the photon energy (1.7--3.1 keV in
red, 3.1--5 keV in green, and 5--8 keV in blue). The bright source
in white is \sedici . The bluish diffuse source is most likely a
cluster of galaxies. The soft X--ray (in red) diffuse emission can
be associated to the SNR G337.0--0.1. \label{fig:ima1627}}
\end{figure}

In May 2008 \sedici\ started a new period of bursting activity.
Its X--ray luminosity reached a level  higher  than that observed
in 1998 (Fig.~\ref{fig:lc1627}) and several short bursts were also
detected. Thanks to a series of Swift observations we could
monitor in detail the outburst evolution~\cite{esp08}. The X--ray
flux initially showed a  rapid decrease, later followed by a
shallower phase consistent with a power law of index $\sim$--0.2.
In Fig.~\ref{fig:decays} the light curve of the 2008 outburst is
compared with that of the previous outburst from this source, and
with the behavior seen after the outbursts of a few other
AXPs/SGRs~\cite{esp08}. This figure shows that, whenever early
data are available, they indicate that a single power law decay
cannot reproduce the source fading. This is due to the presence of
a steeper initial phase in the first days after the outburst and
suggests the presence of two different mechanisms at play. One
possibility is that the steep phase be due to magnetospheric
currents dissipation while the later phase reflect the effect of
crustal cooling. It is also possible that X-rays emitted during
the initial bright burst, delayed by interstellar dust scattering,
contribute to the initial steep phase (see below).

Due to visibility constraints, XMM-Newton could not observe
\sedici\ until September 2008, thus the brightest part of the
outburst could not be observed with this satellite. A Target of
Opportunity observation was performed on 2008 September 27-28, and
despite the low source flux $\sim3\times10^{-13}$ erg cm$^{-2}$
s$^{-1}$, the large effective area of the EPIC instrument allowed
us to collect enough counts to perform a meaningful timing
analysis. This led  to the discovery of  the long-sought
pulsations \cite{esp09}. The spin period is 2.6 s, one of the
shortest among magnetar candidates. The X--ray pulse profile,
characterized by two peaks of different intensity, is shown in
Fig.~\ref{fig:fol1627}.

The deep XMM-Newton observation also showed, for the first time,
diffuse X--ray emission from the vicinity of \sedici
(Fig.~\ref{fig:ima1627}). This consists of two spectrally distinct
components with different spatial extent. The harder emission is a
spatially resolved source located about 1.5 arcmin south of the
SGR. Its high absorption and evidence for a redshifted Fe line
suggest that it might be a  cluster of galaxies. The softer
emission is more extended and most likely related to the supernova
remnant / HII region complex CTB~33~\cite{sar97}.

\subsection{1E 1547.0--5408}

The transient X--ray source \sgr\ was discovered with the
\textit{Einstein Observatory} almost 30 years ago~\cite{lam81} in
the supernova remnant G\,327.24--0.13.  It attracted little
interest until new X-ray and optical studies ruled out more
standard interpretations and led to propose it as an AXP
candidate~\cite{gel07}. This suggestion was confirmed by the
subsequent discovery~\cite{cam07c} of radio pulsations with
$P=2.1$ s and period derivative $\pdot=2.3\times10^{-11}$ s
s$^{-1}$. In October 2008 \sgr\ started an outburst with the
emission of several short bursts and a significant increase in its
X-ray flux~\cite{isr09}.

\begin{figure}
  \includegraphics[height=.3\textheight]{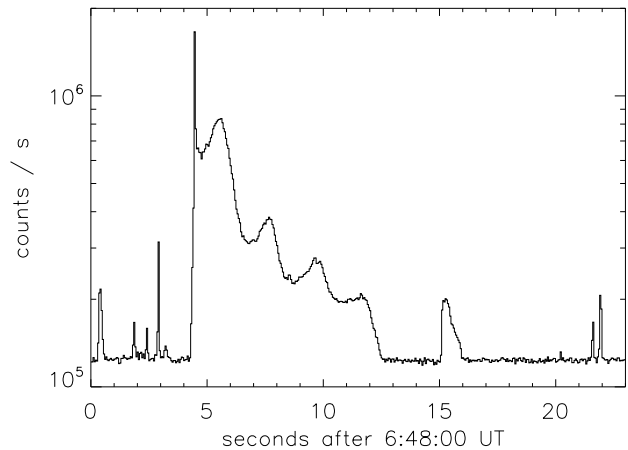}
 \caption{Bursts from \sgr\ observed at E$>$80 keV with the
Anti-Coincidence System of the SPI instrument on board INTEGRAL on
January 22, 2009. The initial spike of the longest burst had a
duration of $\sim$0.3 s and reached a peak flux greater than 2
10$^{-4}$ erg cm$^{-2}$ s$^{-1}$ (25 keV - 2 MeV). A modulation at
2.1 s,  reflecting  the neutron star rotation period,  is clearly
visible in the burst tail.
  \label{fig:burst1547}}
\end{figure}

\begin{figure}
  \includegraphics[angle=270,width=12cm]{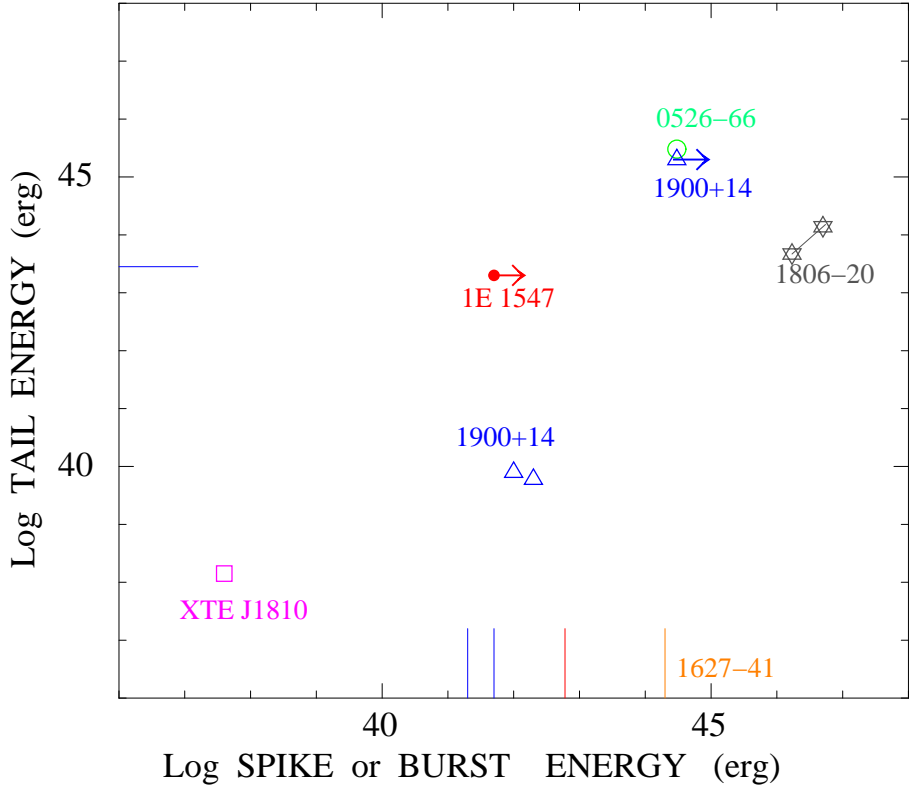}
 \caption{Energetics of flares and peculiar bursts from SGRs and
AXPs. The different sources are distinguished by the symbols
color. The ordinate gives the energy in the pulsating tails that
often follow the brightest bursts, while the abscissa reports the
energy in the initial spikes (data from Mereghetti et al. (2009)
and references therein). The vertical/horizontal lines refer to
events in which only one of these components has been observed.
The three historical giant flares from SGRs are in the upper right
corner. Note that in some cases only lower limits to the total
energy could be derived due to instrument saturation. The two
points for SGR 1806--20 are for the generally assumed distance of
15 kpc and for the more recent estimate d=8.7 kpc. The energetics
of the burst from \sgr\, for an assumed distance of 10 kpc, is in
the range of the so called ``intermediate flares''.
  \label{fig:ene}}
\end{figure}

\begin{figure}
  \includegraphics[height=.3\textheight]{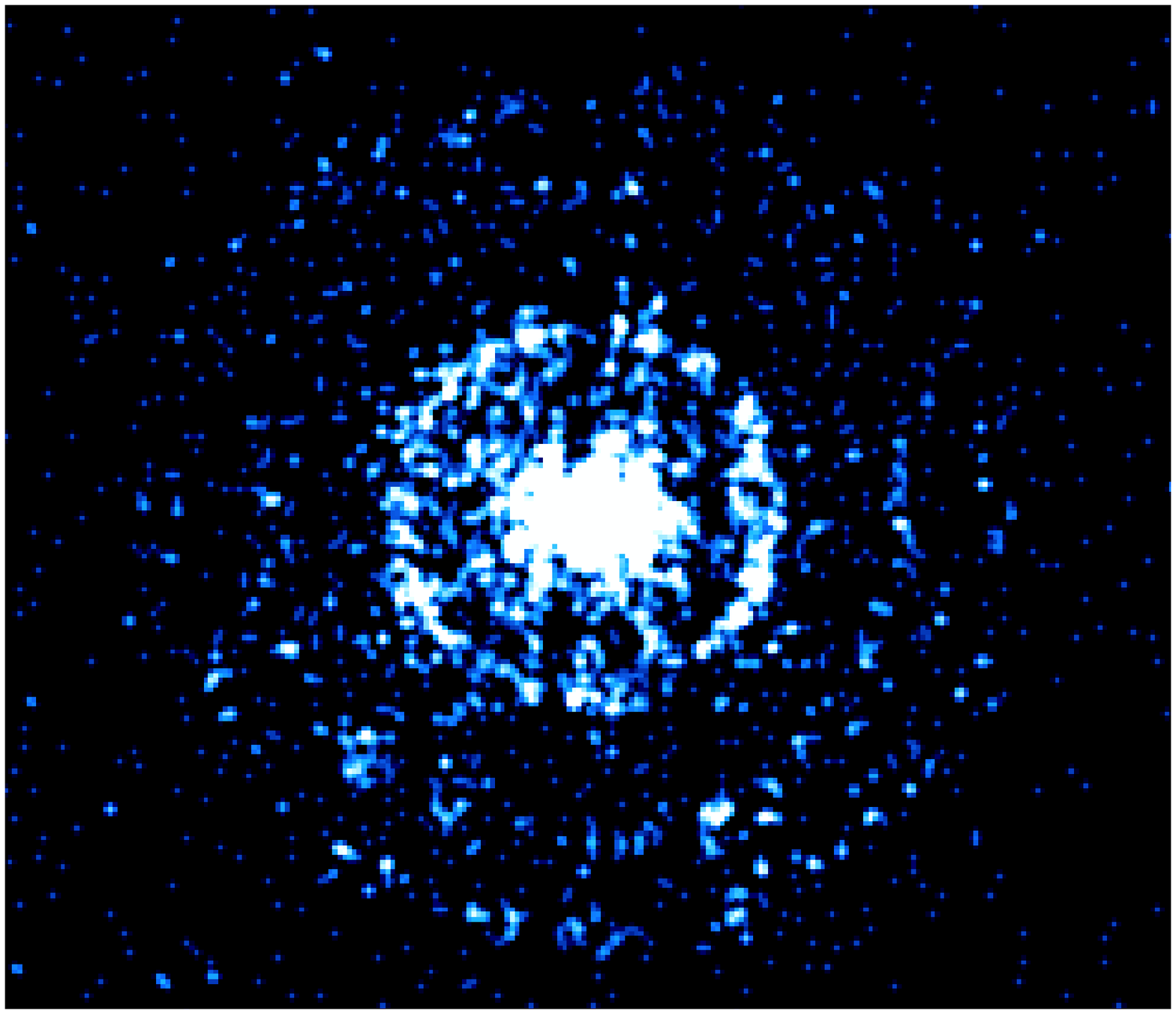}
 \caption{X-ray rings produced by dust scattering around \sgr .
In this image, obtained with the Swift/XRT instrument on January
23, the innermost and brightest ring has a radius of $\sim$1
arcmin. Two outer rings, produced by closer dust layers are also
visible. The ring dimensions were seen to increase in later
observations, as expected for scattering by narrow dust layers of
the X--ray flux emitted during the strong bursting activity that
took place around 6:48 UT of January 22. \label{fig:ring}}
\end{figure}

The bursting activity from \sgr\ culminated on 2009 January 22,
when more than 200 bursts were detected in a few hours. Some of
these bursts were particularly bright, reaching a peak flux above
2$\times$10$^{-4}$ erg cm$^{-2}$ s$^{-1}$ at E$>$25 keV. While
most of the bursts had durations of few hundreds milliseconds, as
typical SGR bursts, two bright events lasted several seconds and
showed a clear modulation at the neutron star spin period. The
light curve of the most interesting one is  displayed in
Fig.~\ref{fig:burst1547}: the burst is composed of a  very bright,
short   ($\sim$0.3 s) initial spike,  followed by a $\sim$8 s long
pulsating tail~\cite{mer09}. These features  are typical of giant
flares from SGRs \cite{maz79,hur99,pal05,mer05b}.  However, the
analysis of INTEGRAL data obtained with the SPI Anticoincidence
Shield indicates that the energy released in this event was only
of the order of a few times 10$^{43}$ erg (for an assumed distance
of 10 kpc), which is  smaller than that of the  three historical
giant flares. The energetics of the strongest bursts and flares
from SGRs/AXPs are compared  in Fig.~\ref{fig:ene}. As mentioned
above, giant flares are characterized by an initial spike followed
by a pulsating tail: the plot  axis refer to these two features.
Note that in some cases only lower limits to the emitted  energy
could be derived due to instrument saturation. The two points for
SGR 1806--20 are for the generally assumed distance of 15 kpc and
for the more recent estimate d=8.7 kpc \cite{bib08}. The values
reported in Fig.~\ref{fig:ene} clearly indicate that there is a
rather continuous distribution of intensities, from the typical
short bursts up to the brightest giant flares.  It is also
noteworthy that extended pulsating tails have been detected not
only for the three giant flares, but also after less intense
bursts~\cite{ibr01,len03,woo05}. Conversely, also a few examples
of pulsating tails apparently without a bright initial hard spike
have been observed~\cite{gui04,mer09}. This is possibly an
indication that the spike emission is non-isotropic, a fact that
adds a some uncertainty to proper estimates of the involved
energy.

Immediately after the discovery of the strong bursting activity of
January 22, several follow-up pointings of \sgr\ were carried out
with the Swift satellite. During the first Swift/XRT observations,
the imaging mode could not be used because the source was too
bright. The first data providing full imaging
(Fig.~\ref{fig:ring}) were obtained on January 23 at $\sim$15:30
UT and showed the presence of remarkable dust scattering rings
around the source position \cite{tie09}. Further observations
carried out with Swift, XMM-Newton and Chandra clearly showed that
the  angular size of the three rings  increased with time. Dust
scattering X-ray halos around bright galactic sources were
predicted well before their observations with the first X-ray
imaging instruments~\cite{ove65}. Their study gives information on
the properties and spatial distribution of the interstellar dust.
When the scattered radiation is a short burst/flare and the dust
is concentrated in a relatively narrow cloud, an expanding ring
(instead of a steady diffuse halo) appears, due to the difference
in path-lengths at different scattering angles. X-ray expanding
rings due to dust scattering have been observed in a few gamma-ray
bursts, and through their study accurate distances of the
scattering dust clouds  in our galaxy could be
determined~\cite{vau04,tie06,via07}.

The dust scattering rings around \sgr\ are the brightest ever
observed and the first ones for an AXP/SGR.  By fitting their
expansion law it is possible to determine the burst emission time,
which is found to coincide with the interval of highest activity
including the bright event of $\sim$6:48 UT shown in
Fig.~\ref{fig:burst1547}. A comprehensive spectral analysis of all
the available X--ray data of the expanding rings around \sgr\ will
allow us to determine the distances of the source and of the three
dust layers~\cite{tie09b}.


\begin{theacknowledgments}
We acknowledge the support from ASI contract I/088/06/0.
\end{theacknowledgments}



\bibliographystyle{aipproc}   


\IfFileExists{\jobname.bbl}{}
 {\typeout{}
  \typeout{******************************************}
  \typeout{** Please run "bibtex \jobname" to optain}
  \typeout{** the bibliography and then re-run LaTeX}
  \typeout{** twice to fix the references!}
  \typeout{******************************************}
  \typeout{}
 }



\end{document}